\documentclass[twocolumn,showpacs,preprintnumbers,superscriptaddress,amsmath,amssymb]{revtex4-2}
\usepackage{amssymb}		
\usepackage{amsmath}
\usepackage{graphicx}
\usepackage{subfigure}
\usepackage[normalem]{ulem}

\usepackage{multirow}
\usepackage{appendix}
\usepackage{float}
\usepackage{CJK}
\usepackage[usenames]{color}
\usepackage{bm}
\usepackage[colorlinks,linkcolor=blue,urlcolor=blue,citecolor=blue]{hyperref}
\usepackage{booktabs} 	     
\usepackage{leftidx}

\usepackage{tikz,xcolor,hyperref}
\definecolor{lime}{HTML}{A6CE39}
\DeclareRobustCommand{\orcidicon}{
	\begin{tikzpicture}
	\draw[lime, fill=lime] (0,0) 
	circle [radius=0.16] 
	node[white] {{\fontfamily{qag}\selectfont \tiny ID}};
	\draw[white, fill=white] (-0.0625,0.095) 
	circle [radius=0.007];
	\end{tikzpicture}
	\hspace{-2mm}
}

\foreach \x in {A, ..., Z}{%
	\expandafter\xdef\csname orcid\x\endcsname{\noexpand\href{https://orcid.org/\csname orcidauthor\x\endcsname}{\noexpand\orcidicon}}
}

\usepackage{soul,xcolor}
\setstcolor{red}
\newcommand\redsout{\bgroup\markoverwith{\textcolor{red}{\rule[0.5ex]{2pt}{0.4pt}}}\ULon}

\usepackage[thicklines]{cancel}
\usepackage{xcolor}

\makeatletter

\newcommand{\Rmnum}[1]{\expandafter\@slowromancap\romannumeral #1@}
\makeatother
\begin{document}
\begin{CJK*} {UTF8} {gbsn}

\title{Collective flows of protons and deuterons in Au + Au collisions at $E_{beam} = 1.23$$A$ GeV by the IQMD model}

\author{Ling-Meng Fang(房灵猛)}
\affiliation{Shanghai Institute of Applied Physics, Chinese Academy of Sciences, Shanghai 201800, China}
\affiliation{University of Chinese Academy of Sciences, Beijing 100049, China}

\author{Yu-Gang Ma(马余刚)\orcidB{}}\thanks{Email:  mayugang@fudan.edu.cn}
\author{Song Zhang(张松)\orcidC{}}\thanks{Email: song\_zhang@fudan.edu.cn}
\affiliation{Key Laboratory of Nuclear Physics and Ion-beam Application (MOE), Institute of Modern Physics, Fudan University, Shanghai 200433, China}
\affiliation{Shanghai Research Center for Theoretical Nuclear Physics， NSFC and Fudan University, Shanghai 200438, China}

\date{\today}
\begin{abstract}
Collective flows of protons and deuterons for Au + Au collisions at beam energy $E_{beam}$ = 1.23$A$ GeV were simulated by an Isospin dependent Quantum Molecular Dynamics (IQMD) model. Two coalescence models, namely naive coalescence  and dynamical coalescence models,  for the formation of deuterons are compared. 
After reasonable match of rapidity spectra of protons and deuterons to the High Acceptance DiElectron Spectrometer (HADES) data is reached，
we apply an event-plane method to calculate the first four-order collective flow coefficients as well as the ratios of $\left \langle v_{4} \right \rangle/\left \langle v_{2} \right \rangle^{2}$ and $\left \langle v_{3} \right \rangle/(\left \langle v_{1} \right \rangle \left \langle v_{2} \right \rangle)$, and observe the number of constituent nucleon scaling among protons and deuterons. In addition, the dependence of $\varepsilon_{n}$ and $\left \langle v_{n} \right \rangle$ as well as the ratio $\left \langle v_{n} \right \rangle/\varepsilon_{n}$ on the centrality is obtained. Lastly, we further investigate the Pearson coefficients $corr(v_{n},v_{m})$  between the first four harmonic flows for protons and deuterons as a function of rapidity and centrality.

 	\end{abstract}
\maketitle

	\section{Introduction}

In heavy-ion collisions, a highly excited nuclear medium is created, and its collective expansion produces the associated particle emission. In a perfectly central collision, the expansion should be isotropic in the transverse plane, as observed in the transverse mass spectra of the ejected particles. The shape of the overlapping regions becomes more anisotropic in  more off-central collisions. In heavy-ion collisions, the collective motion of final-state particles can be described by the collective flows, which can be divided into longitudinal flow and transverse flow according to the motion direction of the final-state particles. The anisotropic flow is essentially originated from  the asymmetrical azimuthal distribution of participant nucleons, which can be classified into directed flow, elliptic flow, triangular flow, quadruple flow and so on  according to different terms of the Fourier expansion of the azimuthal distribution. 

The Fourier expansion  of the azimuthal distribution of the final-state emission particles in momentum space can be expressed as follows~\cite{ollitrault1992anisotropy,poskanzer1998methods,voloshin1996flow}:
\begin{equation}
       E\frac{d^{3}N}{d^{3}p} = \frac{1}{2\pi}\frac{d^{2}N}{p_{t}dp_{t}dy}\left ( 1+\sum_{n=1}^{\infty}2v_{n}\cos\left [ n\left ( \phi -\Psi _{r} \right ) \right ] \right )
       \end{equation}
where $E$ is the energy of the final particle, $p_{t}$ is the transverse momentum of the particle, $y$ is the rapidity, $\phi$ is the azimuthal angle of the transverse momentum relative to the fixed plane $XZ$, $\Psi_{r}$ is the azimuthal angle of the reaction plane relative to the fixed plane $XZ$.  $v_n$ at $n$ = $1, 2, 3, 4 $ are defined as directed flow, elliptic flow, triangular flow,  and quadruple flow,  respectively, as mentioned before.

In general we know that, the elliptic shape of the transverse momentum distribution of the final particles is located in-plane in lower energy below a hundred MeV per nucleon  due to the collective rotation dominated by the attractive mean-field \cite{Wilson,flow_ma,Yan_PLB,WangSS_EPJA,Shi_NST}. With the increasing of beam energy to a few hundred MeV energy, the elliptic shape could be perpendicular to the reaction plane in the mid-rapidity region which is mainly because the spectators have not moved away from the reaction area  timely at such energy range \cite{squeeze,Wang_PRC,WangTT_EPJA}. The spectators have a shadowing effect on participants, making particles tend to eject perpendicular to the direction of the reaction plane. This phenomenon is called "squeeze-out effect", i.e. an elliptical flow outside of the reaction plane. While in the high energy region, because of the Lorentz contraction in two nuclei collisions, the transverse size of the nucleons is negligible relative to the longitudinal alignment. The time for the two nuclei to cross is extremely short in comparison with the characteristic time of elliptic flow formation, thus the bystander leaves the reaction area quickly and almost has no shadowing effect on the reaction zone, so the final particles tend to exude in the reaction plane, and the elliptic flow is in the reaction plane \cite{Heinz}. 

In 1992, Ollitrault {\it et al.} found that the spatial energy density distribution at the early stage of the collision was related to the spatial angular distribution of the freeze-out particles at the later stage of the reaction ~\cite{gyulassy1979pion}. In 1996,  Voloshin {\it et al.}  carried out the Fourier expansion of the particle spectrum of the final state particles, and proposed a method to express the size of the collective flows of the final state particles by the coefficients of the expansion terms ~\cite{voloshin1996flow}. After that, with the continuous in-depth theoretical researches, people studied the collective flows of each order in details, and put forward different calculation methods of collective flows, for example, event-plane method~\cite{danielewicz1985transverse,heinz2002two}, energy momentum tensor method~\cite{ollitrault1993determination} and two particle correlation method~\cite{ollitrault1993determination,ma1995,poskanzer1998methods,borghini2001flow}, etc. With the development of the accelerators, high-energy heavy ion collision experiments can be carried out under different conditions to study the collective flows of final state particles at different energies. In 1999, Heiselberg and Levy studied the azimuthal asymmetry of the system reflected by elliptical flow in noncentral collisions ~\cite{heiselberg1999elliptic}. Stachel concluded that the energy dependence of elliptical flow in high-energy heavy ion collisions is related to QGP phase transition after analyzing the experimental data of several accelerators ~\cite{stachel1999towards}. Voloshin and Poskanzer analyzed Pb + Pb collisions on SPS and found that the elliptic flow has the centrality and rapidity dependence ~\cite{poskanzer1999centrality}. In 2000, Heinz {\it et al.} investigated anisotropic flows and established a deeper connection with QGP phase transitions ~\cite{kolb2000anisotropic}. 

Recently, the HADES Collaboration made systematic measurements on properties of baryon-rich matter formed  in Au + Au collisions at $\sqrt{s_{NN}}$ = 2.4 GeV. Different probes, including dilepton and virtual photons \cite{hades_natphys}, identical pion intensity interferometry \cite{hades_hbt,hades_hbt2} as well  high-order harmonic flows of light nuclei ~\cite{adamczewski2020directed}, which provide an opportunity to investigate the nuclear fireball properties as well as light nuclei production mechanism \cite{Ko,FangLM,Li1_SCPMA,Li2_SCPMA}, and then constrains theoretical model in this reaction energy region and contributes the understanding of the `ice in the fire’ puzzle~\cite{PhysRevC.105.044909}. 

The paper is organized as follows. First, a concise introduction to the IQMD model and coalescence model as well as  the event-plane method for flow analysis are given in Sec.~\ref{sec:model}. Next, the results of first to fourth order coefficients of collective flow of protons and deuterons are presented in Sec.~\ref{sec:analysis}. The results about the linear correlations between different-order flows and eccentricity are also given in this section. Finally, a brief summary is presented in Sec.~\ref{sec:summary}.

	\section{Models and methods}
	\label{sec:model}

In the study of heavy-ion collisions, various models have been established to simulate the collision processes. At present, the commonly used heavy-ion reaction models can be divided into statistical models and transport models.

In this study, an Isospin dependent Quantum Molecular Dynamics (IQMD) model, a kind of transport models, is employed to study the reaction system from initial state to final stage in medium-high energy heavy-ion collisions. The coalescence model is used to simulate the generation of light nuclei  by using the nucleon phase-space from IQMD model. And the collective flow of light nuclei are calculated from the phase-space information at the freeze-out stage simulated by the IQMD model with help of the event-plane method. In the following, the IQMD model, coalescence model and event plane calculation methods will be introduced, separately.

	\subsection{The IQMD model}
 
Quantum Molecular Dynamics (QMD) model can provide the information on both the collision dynamics and the phase space information~\cite{aichelin1991quantum,ma1995onset,feng2018nuclear,zhang2021hyperon,guo2020secondary}. The IQMD model is based on the traditional QMD model, by including the isospin degree of freedom of nucleons~\cite{hartnack1998modelling}.

In the IQMD model, the normalized wave function of each nucleon is expressed in the form of a Gaussian wave packet, 
      \begin{equation}
      \phi_{i}(\vec{r},t) = \frac{1}{(2\pi L)^{3/4}}\exp(\frac{-(\vec{r}-\vec{r_{i}}(t))^{2}}{4L})\exp(\frac{i\vec{r}\cdot \vec{p_{i}}(t)}{\hbar}),
      \end{equation}
here $\vec{r_{i}}(t)$ and $\vec{p_{i}}(t)$ are time-dependent variables describing the center of the wave packet in coordinate space and momentum space, respectively. Given the direction of $\vec{r_{i}}$ and $\vec{p_{i}}$, $\phi_{i}(\vec{r},t$) is a four-dimensional function. The parameter $L$ is the width of the wave packet, which is related to the size of the reaction system and usually fixed in the simulations. Here the width $L$ is fixed as $2.16fm^{2}$ for Au + Au reactions~\cite{liu2017mean,yu2020investigation}.

All the nucleons interact with each other through an effective mean field and two-body scatterings. The interaction potential can be expressed as
      \begin{equation}
      U = U_{Sky} + U_{Coul} + U_{Yuk} + U_{Sym} + U_{MDI},
      \end{equation}
where $U_{Sky}$, $U_{Coul}$, $U_{Yuk}$, $U_{Sym}$, and $U_{MDI}$ represent the density-dependent Skyrme potential, Coulomb potential, Yukawa potential, isospin asymmetric potential, and the momentum-dependent interaction potential, respectively. The nucleon-nucleon collision cross section in the medium ($\sigma_{NN}^{med}$) can be expressed as taken in Refs.~\cite{westfall1993mass,chen1968vegas,zhang2020probing}
      \begin{equation}
      \sigma_{NN}^{med} = (1-\eta \frac{\rho}{\rho_{0}})\sigma_{NN}^{free},
      \end{equation}
where $\rho_{0}$ is the density of normal nuclear matter, $\rho$ is the local density, $\eta$ is the in-medium correction factor, which is chosen as 0.2 in this paper to better reproduce the flow data~\cite{wang2018effect}, and $\sigma_{NN}^{free}$ is the free nucleon-nucleon cross section.

      \subsection{Coalescence model}
      
There are two types of coalescence models, naive coalescence model and dynamical coalescence model. In this article, we use both of two coalescence models, and compare the difference between them.

The naive coalescence model uses the following criteria to judge the formation of deuterons:
      \begin{equation}
      \Delta p<p_{0}, ~~~~~ \Delta r<r_{0},
      \end{equation}
where $\Delta p = \left |\vec{p_{1}}-\vec{p_{2}} \right |$, $\Delta r = \left |\vec{r_{1}}-\vec{r_{2}} \right |$, and $p_{0} = 0.35$ GeV/$c$, $r_{0} = 3.5$ $fm$ are selected in this paper. It is emphasized that here  the momentum and coordinate should be at the rest frame of the pair, such as proton and neutron.

The dynamical coalescence model can give the probability of light nuclei by the overlap of the cluster Wigner phase-space density with the nucleon phase space distributions at an equal time in the $M$-nucleon rest frame at the freeze-out stage~\cite{chen2003light}. The momentum distribution of a cluster in a system containing $A$ nucleons can be expressed by:
      \begin{eqnarray}
	 \begin{aligned}
      &\frac{d^{3}N_{M}}{d^{3}K}=G\binom{A}{M}\binom{M}{Z}\frac{1}{A^{M}}\int \left [ \prod_{i=1}^{Z}f_{p}\left ( \vec{r}_{i},\vec{k}_{i} \right ) \right ]\\
      &\qquad\left [ \prod_{i=Z+1}^{M}f_{n}\left ( \vec{r}_{i},\vec{k}_{i} \right ) \right ]\times \rho ^{W}\left ( \vec{r}_{i_{1}},\vec{k}_{i_{1}},\cdots ,\vec{r}_{i_{M-1}},\vec{k}_{i_{M-1}} \right )\\
      &\qquad\times \delta \left ( \vec{K}-\left ( \vec{k}_{1}+\cdots +\vec{k}_{M} \right ) \right )d\vec{r}_{1}d\vec{k}_{1}\cdots d\vec{r}_{M}d\vec{k}_{M},
       \end{aligned}
      \end{eqnarray}
where $M$ and $Z$ are the number of the nucleon and proton of the cluster, respectively; $f_{n}$ and $f_{p}$ are the neutron and proton phase-space distribution functions at freeze-out, respectively; $\rho_{W}$ is the Wigner density function; $\vec{r}_{i_{1}}\cdots\vec{r}_{i_{M-1}}$ and $\vec{k}_{i_{1}}\cdots\vec{k}_{i_{M-1}}$ are the relative coordinate and momentum in the $M$-nucleon rest frame; the spin-isospin statistical factor $G$ is $3/4$ for deuteron in this paper~\cite{chen2003light,zhao2018spectra,sun2021light}. While the neutron and proton phase-space distribution comes from the transport model simulations, the multiplicity of a $M$-nucleon cluster is then given by:
      \begin{eqnarray}
	 \begin{aligned}
      &N_{M}=G\int \sum_{i_{1}>i_{2}>\cdots >i_{M} }d\vec{r}_{i_{1}}d\vec{k}_{i_{1}}\cdots d\vec{r}_{i_{M-1}}d\vec{k}_{i_{M-1}}\\
      &\qquad\left \langle \rho _{i}^{W}\left ( \vec{r}_{i_{1}},\vec{k}_{i_{1}},\cdots,\vec{r}_{i_{M-1}},\vec{k}_{i_{M-1}}  \right ) \right \rangle,
      \end{aligned}
      \end{eqnarray}
 where the $\left \langle \cdots \right \rangle$ denotes the event averaging.

      \subsection{The event-plane method  for flow analysis}
      
A common method for calculating collective flow is the event-plane method. The $n$-th order event-plane angle $\Psi _{EP}^{(n)}$ can be defined by the event flow vector $Q_{n,x}$ and $Q_{n,y}$ as~\cite{poskanzer1998methods,adamczyk2013third,ollitrault1992anisotropy,poskanzer1998methods,afanasiev2009systematic,adamczyk2012inclusive}:
      \begin{eqnarray}
	 \begin{aligned}
      &\Psi _{EP}^{(n)} = \frac{1}{n}\tan^{-1}\left ( \frac{Q_{n,y}}{Q_{n,x}} \right ), \\
      &Q_{n,x}= \sum_{i}^{}\omega_{i}\cos(n\Phi_{i}),\quad Q_{n,y}= \sum_{i}^{}\omega_{i}\sin(n\Phi_{i}),
       \end{aligned}
      \end{eqnarray}
where $\Phi_{i}$ and $\omega _{i}$ are the azimuthal angle of the momentum and the weight for the $i$-th particle, respectively. $\omega _{i}$ is usually set to unit for theoretical simulation but set as charges $|Z|$ in this paper, which is suggested in Ref.~\cite{adamczewski2020directed}. The sums extend over all particles used in the event plane reconstruction. For systems with finite multiplicity, the harmonic flow coefficients can be calculated by:
      \begin{eqnarray}
	 \begin{aligned}
      &\left \langle v_{n}\right \rangle=\frac{\left \langle v_{n}^{obs}\right \rangle}{Res\left \{ \Psi _{n}\left \{ EP \right \} \right \}},\\
      &\left \langle v_{n}^{obs}\right \rangle=\left \langle \cos\left ( km\left ( \phi -\Psi _{n}\left \{ EP \right \} \right ) \right ) \right \rangle,\\
      &Res\left \{ \Psi _{n}\left \{ EP \right \} \right \}=\left \langle \cos\left ( km\left ( \Psi _{n}\left \{ EP \right \}-\Psi _{RP} \right )\right ) \right \rangle.
       \end{aligned}
      \end{eqnarray}

The angular brackets indicate an average over all particles in all events and $km$ = $n$ in this work. The resolution of event plane angle $Res\left \{ \Psi _{n}\left \{ EP \right \} \right \}$ owing to the finite number of particles can be calculated by:
      \begin{eqnarray}
	 \begin{aligned}
      &Res\left \{ \Psi _{n}\left \{ EP \right \} \right \}=\left \langle \cos\left ( km\left ( \Psi _{n}\left \{ EP \right \}-\Psi _{RP} \right )\right ) \right \rangle\\
      &=\frac{\sqrt{\pi }}{2\sqrt{2}}\chi _{m}\exp\left ( -\chi _{m}^{2}/4 \right )\\
      &\times \left [ I_{\left ( k-1 \right )/2}\left ( \chi _{m}^{2}/4 \right )+ I_{\left ( k+1 \right )/2}\left ( \chi _{m}^{2}/4 \right )\right ],
       \end{aligned}
      \end{eqnarray}
where the $\chi_{m}$ can be estimated by the sub-event method. The event used to calculate the event plane angle would randomly be split into two sub-events, event $A$ and $B$, with maximum difference of particle number equal to 1. $\chi_{m}$ from sub-event resolution $\cos(km(\Psi_{m}^{A}-\Psi_{m}^{B}))$ multiplying $\sqrt{2}$ would be the $\chi_{m}$ for full event resolution $Res\left \{ \Psi _{n}\left \{ EP \right \} \right \}$. The details for this analysis can be found in Refs.~\cite{adamczyk2013third,poskanzer1998methods,afanasiev2009systematic,adamczyk2012inclusive,Zhang:2018zzu}

      \section{Analysis and discussion}
      \label{sec:analysis}

 In this paper, we use an IQMD model to simulate Au + Au collisions at beam energy $E_{beam}$ = 1.23$A$ GeV, which corresponds to a center of mass energy $\sqrt{s_{NN}}$ = 2.4 GeV. The total number of events included in the simulation is 1,600,000. The centrality is characterized as $c = (\pi b^{2})/(\pi b_{max}^{2})\times100\%$, where $b$ is the impact parameter, and $b_{max} = 1.15(A_{P}^{1/3} + A_{T}^{1/3})$ is the sum of effective shape radius of projectile and target. With this definition of centrality, the smaller the $c$ value, the more central the collisions.

   \begin{figure}[htb]
      \includegraphics[angle=0,scale=0.35]{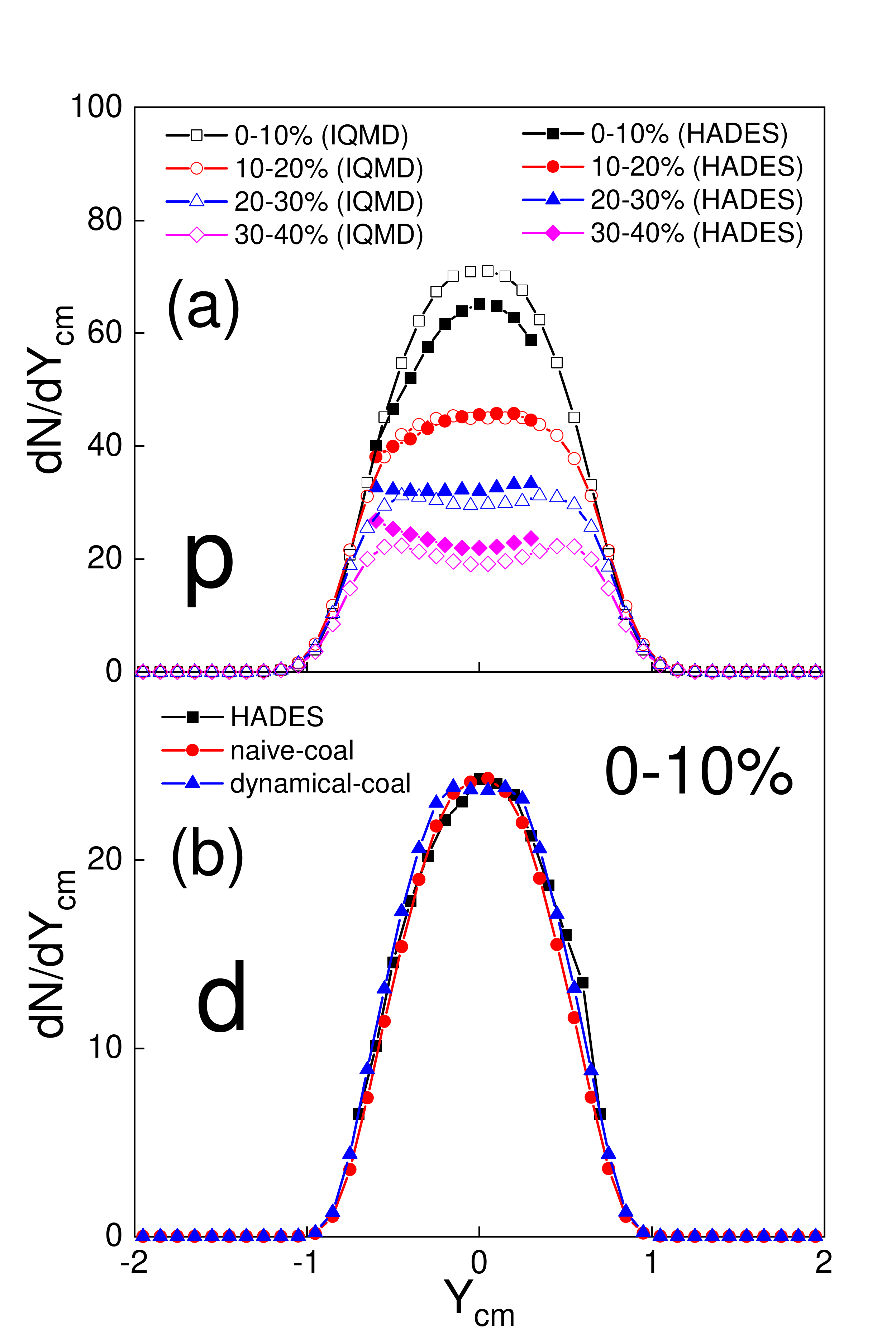}
      \caption{Rapidity distribution of protons (a) and deuterons (b) in Au + Au collisions at $E_{beam}$ = 1.23$A$ GeV for 4 centrality classes with 10\% bin width.  The solid and hollow points in (a) represent the HADES experimental data~\cite{Schuldes:2016eqz} and the IQMD simulation results, respectively. The points in (b) are from the HADES experimental data (black)~\cite{hadesconference}, naive coalescence model (red), and dynamical coalescence model (blue), respectively.}
      \label{fig1}
      \end{figure}
      
\begin{figure}[htb]
      \includegraphics[angle=0,scale=0.35]{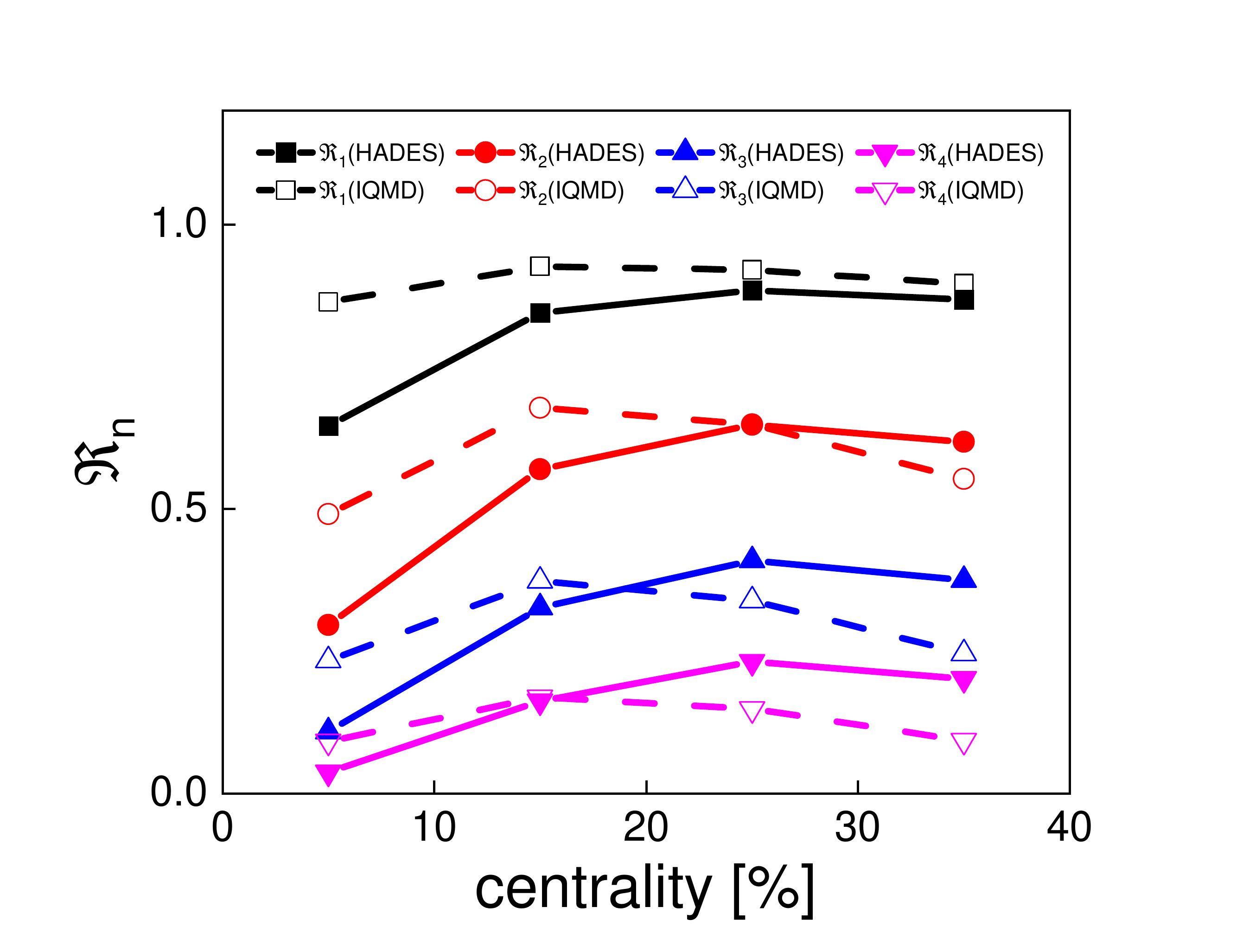}
      \caption{The variation of first to fourth order resolutions with the centrality  in Au + Au collisions at $E_{beam}$ = 1.23$A$ GeV. The solid and hollow points represent the HADES experimental data and the IQMD simulation results, respectively.}
      \label{fig2}
      \end{figure}

    \begin{figure}[htb]
      \includegraphics[angle=0,scale=0.16]{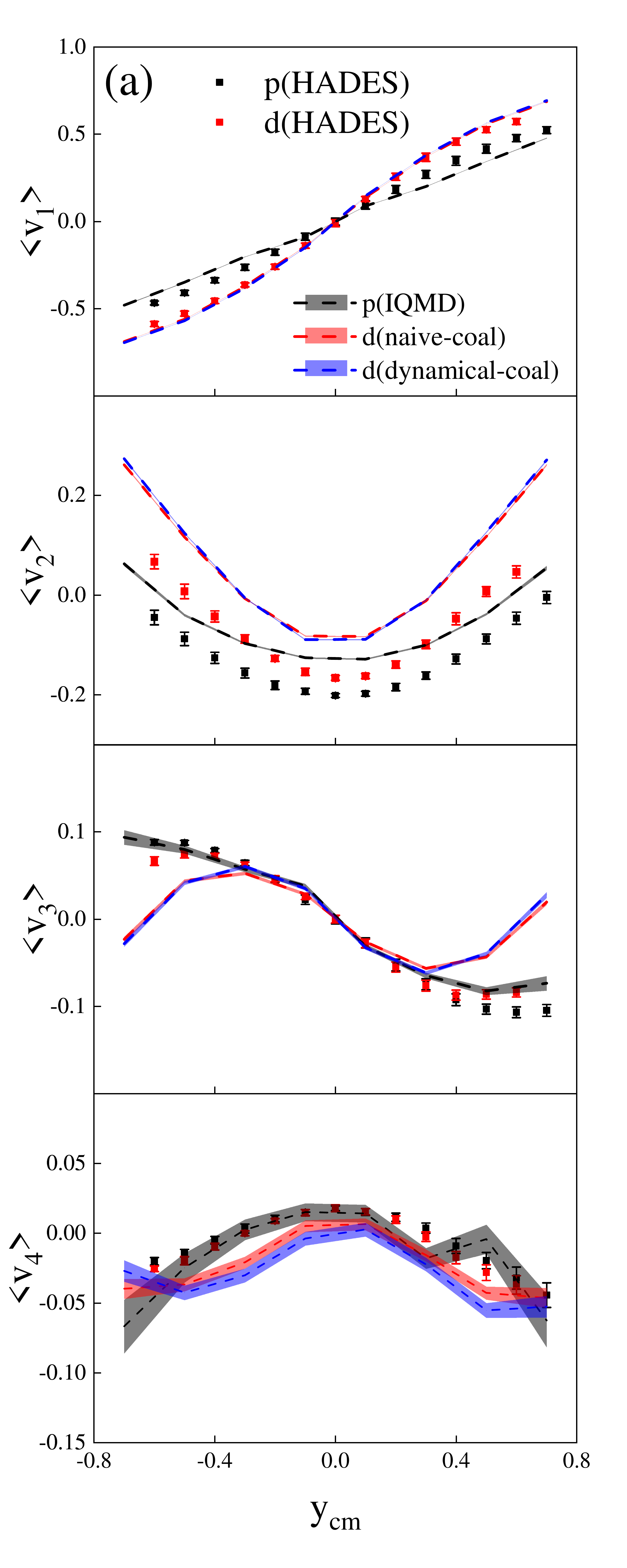}
      \includegraphics[angle=0,scale=0.16]{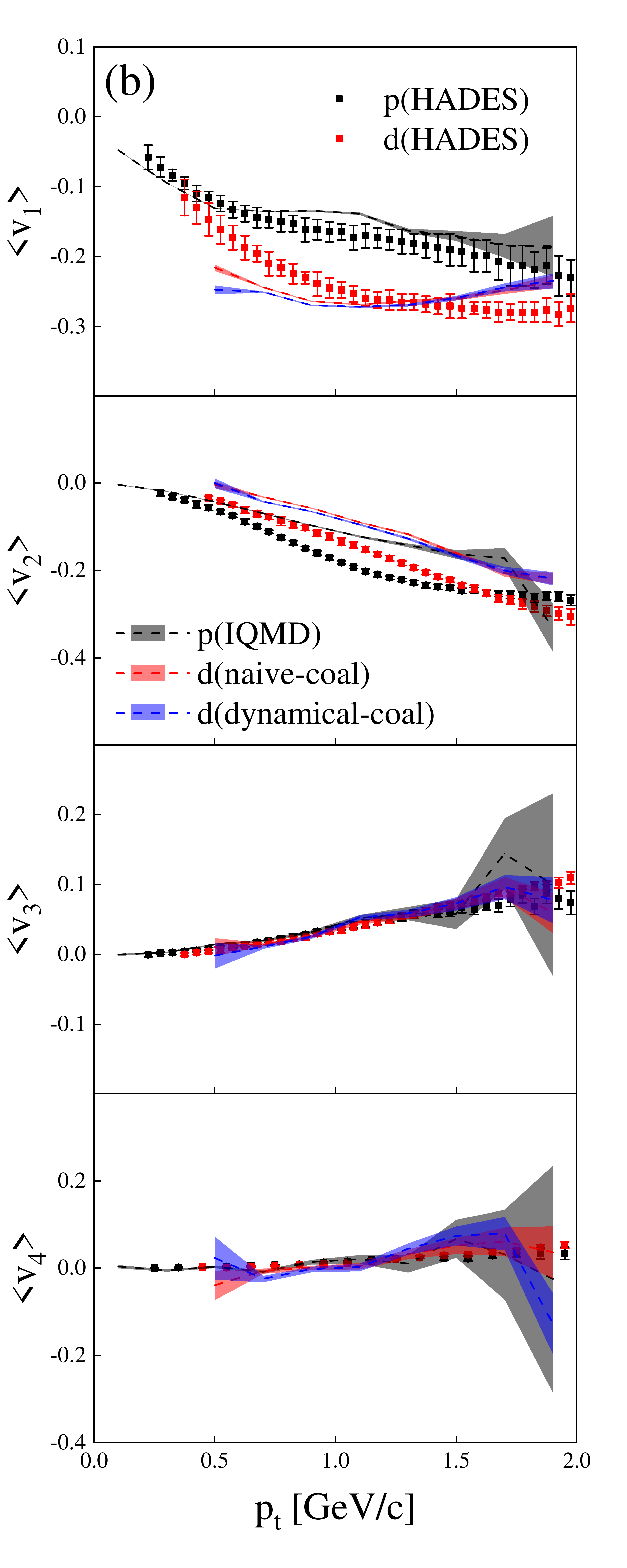}
      \caption{(a) Different harmonic flows as a function of rapidity in Au + Au collisions at $E_{beam}$ = 1.23$A$ GeV for 20-30\% centrality. The protons and deuterons are selected within transverse momentum of  1 - 1.5 GeV/$c$. 
      (b) Different harmonic flows as a function of transverse momentum in Au + Au collisions at $E_{beam}$ = 1.23$A$ GeV for 20-30\% centrality. For the odd-order collective flows, the protons and deuterons are selected within rapidity of -0.25~-0.15, while for the even-order collective flows, the rapidity is selected within -0.05~0.05. In the figure, the solid symbols with error bars represent the HADES experimental data, the dotted lines with error bands represent the IQMD simulation results, respectively.}
      \label{fig3}
      \end{figure}

      \subsection{Yield of protons and deuterons}
      
In this paper, we use naive coalescence model and dynamical coalescence model to estimate deuterons formations. Figure~\ref{fig1} shows the rapidity distributions of protons and deuterons for the 0-10\% centralities as well as the  comparison with the HADES results. 

It is seen from Fig.~\ref{fig1}(b) that the yields of deuterons from two coalescence models are consistent  with each other and are all in good agreement with the HADES experimental data. We notice that the yield of protons from the IQMD model is higher than experimental data in the most central collisions from Fig.~\ref{fig1}(a), but there is an overtaking in more off-centralities. This behavior reproduces previous IQMD results or other models as given in Ref.~\cite{hadesconference}.

      \subsection{Collective flows of protons and deuterons }

To be consistent with the method used by the HADES experiment in Ref.~\cite{adamczewski2020directed}, we use the charges $|Z|$ as the weight in this paper, and the flow coefficients of all orders discussed here are defined relative to $\Psi_{EP,1}$ as:
      \begin{eqnarray}
	 \begin{aligned}
      &\left \langle v_{n}^{obs} \right \rangle = \left \langle \cos\left [ n\left ( \phi -\Psi _{EP,1} \right ) \right ] \right \rangle\\
      &\mathfrak{R}_{n} = \left \langle \cos\left [ n\left ( \Psi _{EP,1} -\Psi _{RP} \right ) \right ] \right \rangle\\
      &\left \langle v_{n} \right \rangle = \left \langle v_{n}^{obs}\right \rangle/\mathfrak{R}_{n}.
      \label{eq:flow}
      \end{aligned}
      \end{eqnarray}

Via this method, we obtain the variation of first to fourth order resolutions versus centrality as shown in Fig.~\ref{fig2}. As we can see from Fig.~\ref{fig2}, the value of resolution decreases significantly as the order increasing, and the resolution obtained by the event-plane method has basically the same trend as the HADES experimental data. In the most central collisions, the emission particles tend to be more isotropic, so the values of all order resolutions are the smallest. With the increase of the centrality value (i.e. more off-central collision), the anisotropy of the emission particles is gradually obvious, so the value of resolution tends to increase gradually. As we can see from Fig.~\ref{fig2}, in the most central collisions, the resolution of IQMD model is higher than that from the HADES, which corresponds to the higher proton yield from IQMD model in the most central collision as shown in Fig.~\ref{fig1}. With the increase of centrality value, the proton yield from the IQMD is gradually lower than that from the HADES, which explains the overtaking phenomenon of resolution in more off-central collisions in Fig.~\ref{fig2}.

Using the event-plane method as in Eq.~(\ref{eq:flow}), we \redsout{can} calculate the distribution of the collective flow as a function of rapidity for light nuclei, as shown in Fig.~\ref{fig3}(a). As we can see from Fig.~\ref{fig3}(a), $\left \langle v_{1} \right \rangle$ and $\left \langle v_{3} \right \rangle$ are anti-symmetric with rapidity, while $\left \langle v_{2} \right \rangle$ and $\left \langle v_{4} \right \rangle$ are axis-symmetric. As for $\left \langle v_{2} \right \rangle$, a negative value in middle rapidity region indicates an out-of-plane emission, which is caused by the so-called squeeze-out effect, where particles are blocked from being emitted in the reaction plane by the spectator nucleons and are therefore emitted mainly in the out-of-plane-direction. As rapidity increases, the value of $\left \langle v_{2} \right \rangle$ becomes positive due to the reduced shadowing effect of bystanders on the reaction zone. And in the middle rapidity region, the $\left \langle v_{2} \right \rangle$ of the protons is lower than that of deuterons, which indicates that after the collision, the protons are more likely to eject out of the plane, while deuterons prefer an in-plane emission. Also, $\left \langle v_{n} \right \rangle$ has a larger magnitude for lower harmonics than higher harmonics. Moreover, we can see that the result of collective flow obtained by the IQMD model is lower than that from the HADES experiment especially for the elliptic flow, this phenomenon is consistent with the results from the UrQMD model in Ref.~\cite{Hillmann:2020wzq,Hillmann:2019wlt}.

The distributions of different order collective flows as a function of light nuclei transverse momentum are shown in Fig.~\ref{fig3}(b). The collective flow  coefficients of deuterons follow that of the free protons, and show a similar strong dependence on the transverse momentum. And the absolute values of collective flow of each order increase with transverse momentum, which indicates that light nuclei with higher $p_{t}$ tend to emit more out of plane, as they are from earlier emission.  $\left \langle v_{1} \right \rangle$ of the deuterons have larger negative values than the protons  which can be inferred from the coalescence mechanism. We can find that the IQMD model can well describe the experimental results of $\left \langle v_{1} \right \rangle$, $\left \langle v_{3} \right \rangle$ and $\left \langle v_{4} \right \rangle$, but $\left \langle v_{2} \right \rangle$ obtained by the IQMD model is slightly lower than that from the HADES experiment.

\begin{figure}[htb]
      \includegraphics[angle=0,scale=0.23]{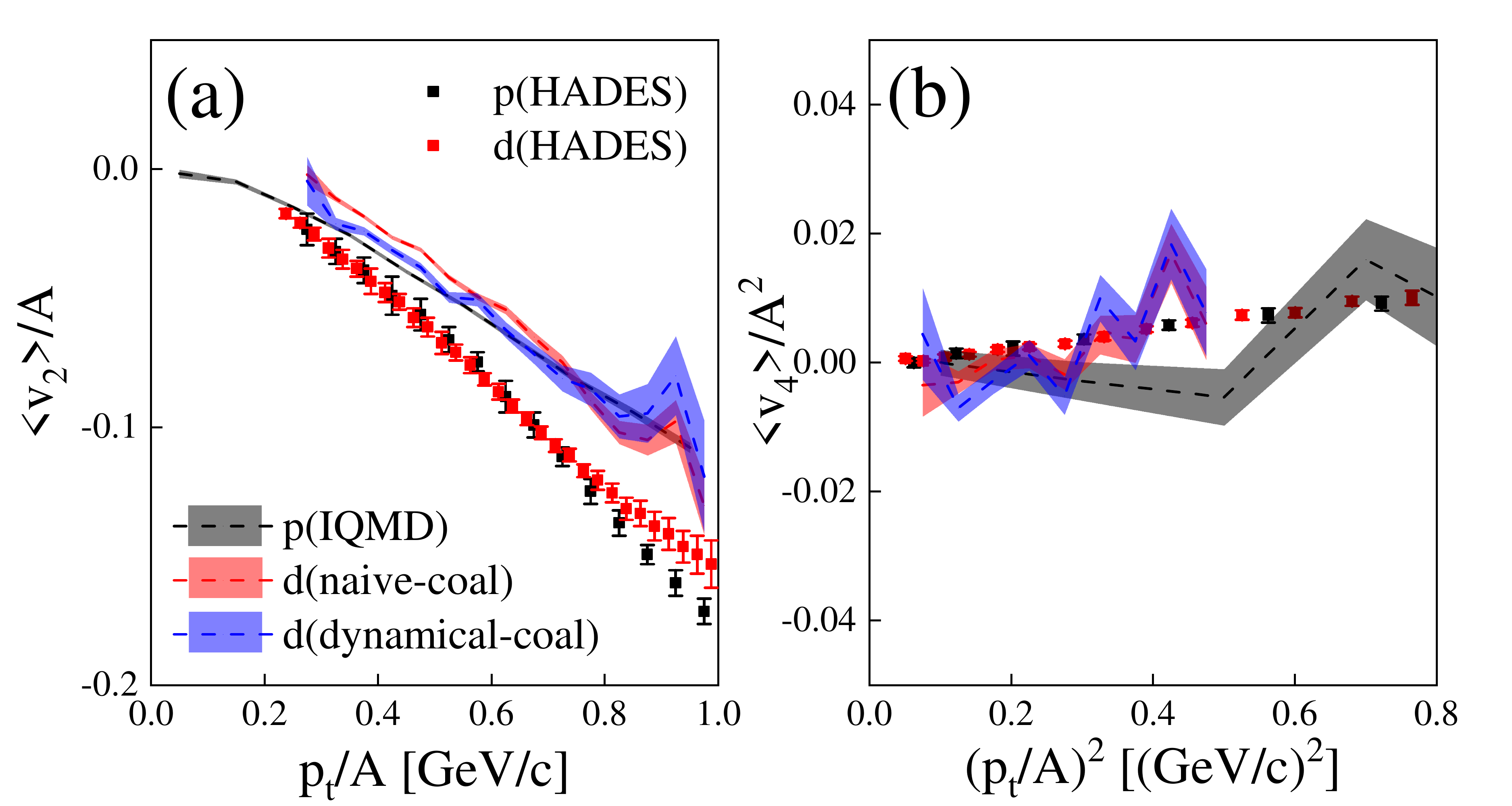}
      \caption{Mass number $A$ scaled $\left \langle v_{2} \right \rangle$(a) and $\left \langle v_{4} \right \rangle$(b) of protons and deuterons  for 1.23$A$ GeV Au + Au collisions in 20-30\% centrality  as a function of transverse momentum per nucleon for $|y|<0.05$.}
      \label{fig4}
      \end{figure}

The scaling of elliptic flow of hadrons with the number of constituents has been established for more than a decade with quark recombination \cite{fries2003hadron} or quark coalescence model~\cite{kolb2004momentum} at RHIC energies,  and an empirical function can also fits the experimental elliptic flow data \cite{Wang_NST}. For the coalescence of nucleons into deuterons the same scaling should be there in terms of the baryon number. It has been first claimed that nucleon-number scaled flows should be observed if the coalescence mechanism is satisfied for light nuclei production in Ref.~\cite{Yan_PLB,Ma_NPA} and later on the experimental confirmation has been achieved by the STAR Collaboration \cite{STAR1}. The nucleon-number scaling of   elliptic flow results in the expectation that $\left \langle v_{2}^{d} \right \rangle\left ( p_{T}^{d} \right ) = 2\left \langle v_{2}^{p} \right \rangle\left ( \frac{1}{2}p_{T}^{d} \right )$. Thus $\left \langle v_{2} \right \rangle/A$ as a function of $p_{T}/A$, with $A$ being the baryon number, should yield the same curves for protons and   light nuclei in the coalescence picture. Moreover, instead scaling by the baryon number $A$ for $\left \langle v_{2} \right \rangle$, the measured data $\left \langle v_{4} \right \rangle$ seems to be scaled by $A^{2}$ in previous studies \cite{Hillmann:2020wzq,Ma_NPA,adamczewski2020directed,Yan:2007vk}. Taking the data of Fig.~\ref{fig3} we show that the flow of protons and the scaled  deuterons for Au + Au collisions in 20-30\% centrality at a beam energy of 1.23 AGeV in Fig.~\ref{fig4}. From Fig.~\ref{fig4}(a) we observe that the simulation predicts a good scaling among protons and deuterons. Fig.~\ref{fig4}(b) display $\left \langle v_{4} \right \rangle/A^{2}$ as a function of $(p_{t}/A)^{2}$, from which we can see that the $\left \langle v_{4} \right \rangle$ can still be roughly scaled by $A^{2}$ for protons and deuterons. However, the scaling behavior is not perfect within the present statistics. For example, the $\left \langle v_{2} \right \rangle/A$ from the IQMD model has a lower magnitude than that from the HADES, which is probably due to the underestimate of $\left \langle v_{2} \right \rangle$ as shown in Fig.~\ref{fig3}. 
      
\begin{figure}[htb]
      \includegraphics[angle=0,scale=0.35]{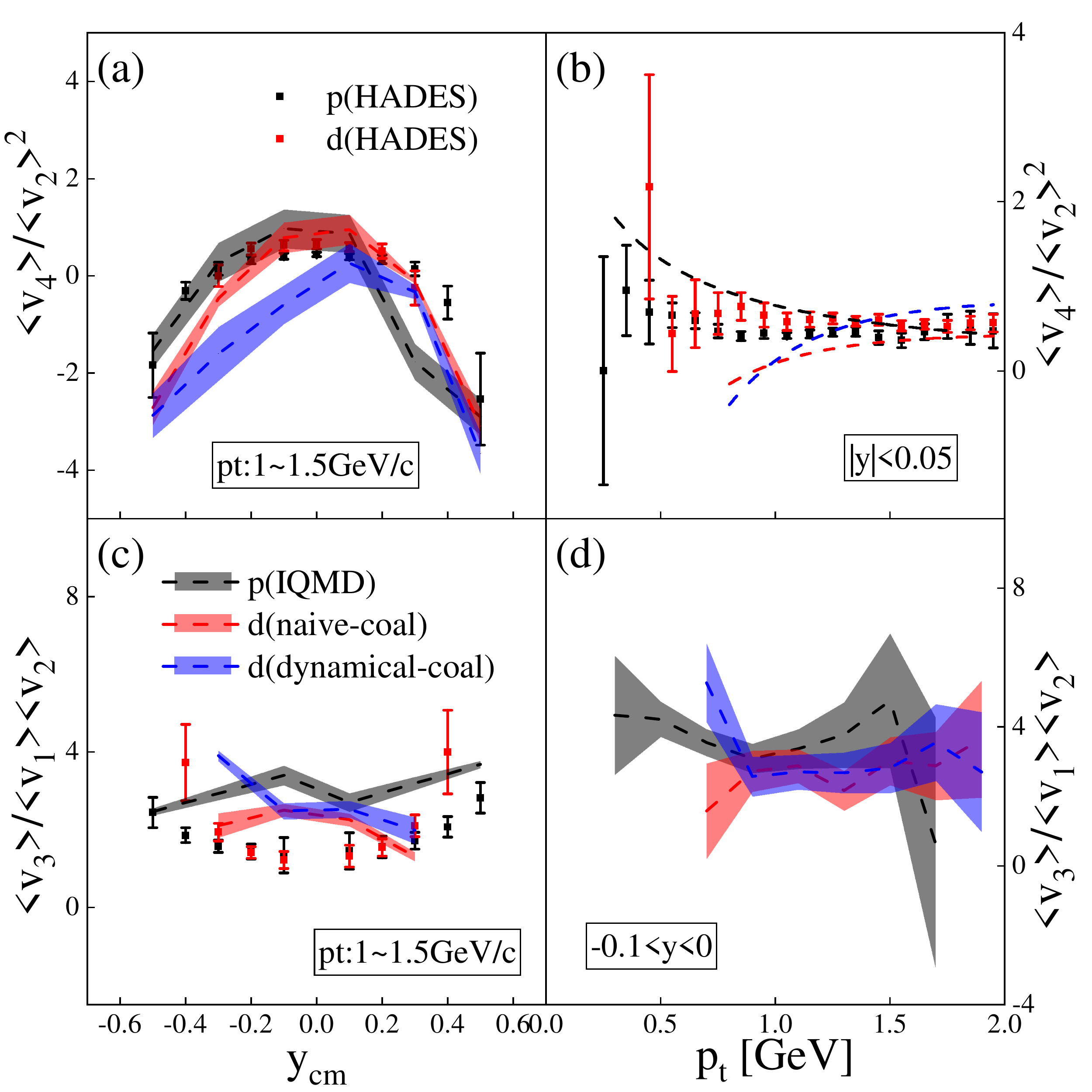}
      \caption{The ratio $\left \langle v_{4} \right \rangle/\left \langle v_{2} \right \rangle^{2}$ (upper row) and $\left \langle v_{3} \right \rangle/(\left \langle v_{1} \right \rangle \left \langle v_{2} \right \rangle)$ (bottom row) distributions on rapidity (left column) and transverse momentum (right column) for protons and deuterons of 1.23A GeV Au+Au collisions at 20-30\% centrality.}
      \label{fig5}
      \end{figure}

The initial fluctuation can affect the initial geometric asymmetry of the overlapping region which could be transferred into   the momentum space partially, and then significantly contribute to  higher-order harmonic flows \cite{Han_PRC}. 
In earlier studies in intermediate energy \cite{Yan_PLB,Ma_NPA} as well as  at ultra-relativistic energies \cite{Han_PRC}, it was found that triangular and quadrangular flows also roughly present a constituent nucleon number scaling in the intermediate-$p_T$ region, similar to the behaviors of elliptic flow. From those results, a nucleon-number  scaling of $\left \langle v_{n} \right \rangle/n^{n/2}$ for different light nuclei holds for harmonic flow ($\left \langle v_{n} \right \rangle$, $n$ = 2, 3, and 4), which can be related to $\left \langle v_{n} \right \rangle$ scaling. In ultra-relativistic energies, such extended flow scaling for high-order harmonic flows has been demonstrated  by the PHENIX Collaboration \cite{Phenix_16} and  STAR Collaboration \cite{STAR_18,STAR_22}. Fig.~\ref{fig5}  show the ratio $\left \langle v_{4} \right \rangle/\left \langle v_{2} \right \rangle^{2}$ and $\left \langle v_{3} \right \rangle/(\left \langle v_{1} \right \rangle \left \langle v_{2} \right \rangle)$ distributions on rapidity and transverse momentum \cite{Kolb_PRC}. As we can see from Fig.~\ref{fig5}(a), for protons and deuterons, the $\left \langle v_{4} \right \rangle/\left \langle v_{2} \right \rangle^{2}$ value approaches to the experimental data of 0.5 within the larger error in mid-rapidity region. However, in the off-middle rapidity interval, the $\left \langle v_{4} \right \rangle/\left \langle v_{2} \right \rangle^{2}$ of protons and deuterons decreases. Fig.~\ref{fig5}(b) demonstrates that the asymptotic values of $\left \langle v_{4} \right \rangle/\left \langle v_{2} \right \rangle^{2}$ of protons and deuterons (naive or dynamical) approach 0.42 and  0.41 or 0.78, respectively, which is overall in agreement with the experimental values. 
As for $\left \langle v_{3} \right \rangle/(\left \langle v_{1} \right \rangle \left \langle v_{2} \right \rangle)$, the results obtained by the IQMD model are higher than those from the HADES experiment, and all of them do not show a significant rapidity correlation.

To  quantify initial geometric asymmetry and fluctuation, an eccentricity is introduced to describe the geometric anisotropy of the overlapping region at the initial state. After considering the initial fluctuation, the eccentricity under the center-of-mass frame is defined as~\cite{alver2010collision,qiu2011event,petersen2012systematic}:
      \begin{eqnarray}
	 \begin{aligned}
      \varepsilon _{n}=\frac{\sqrt{\left \langle r^{n}\cos\left ( n\varphi  \right ) \right \rangle^{2}+\left \langle r^{n}\sin\left ( n\varphi  \right ) \right \rangle^{2}}}{\left \langle r^{n} \right \rangle},
       \end{aligned}
      \end{eqnarray}
      where $r = \sqrt{x^2+y^2}$ and $\varphi$ are the coordinate position and azimuthal angle of participants in the reaction zone with $\langle x\rangle$ = 0 and $\langle y\rangle$ = 0.

\begin{figure}[htb]
      \includegraphics[angle=0,scale=0.23]{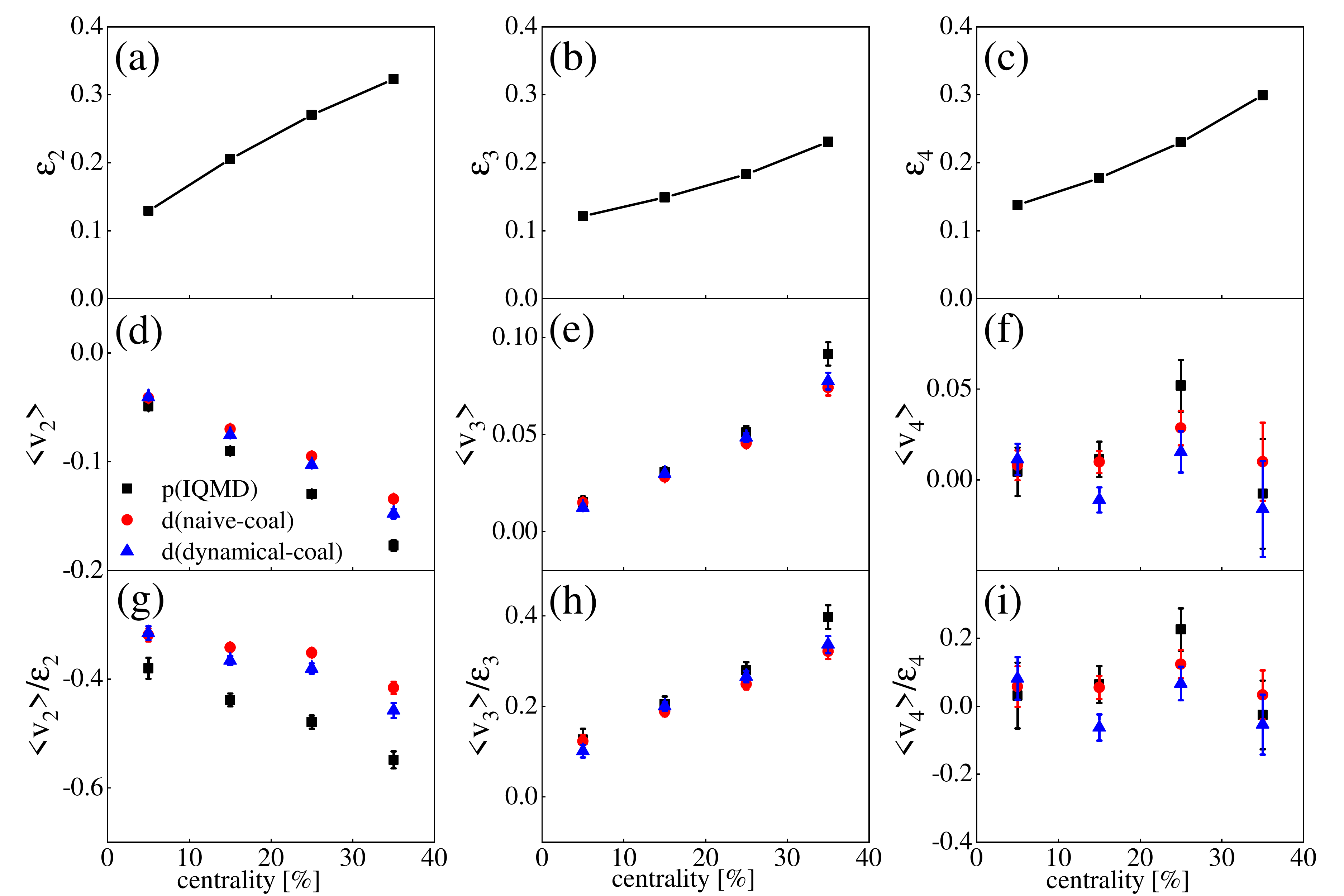}
      \caption{The dependence of the $\varepsilon_{n}$ (top row) and $\left \langle v_{n} \right \rangle$ (middle row) as well as the ratio $\left \langle v_{n} \right \rangle/\varepsilon_{n}$ (bottom row) on the centrality for $n$ = 2 (left column), 3 (middle column), 4 (right column) in Au + Au collisions at $E_{beam} = 1.23A GeV$ for 0-40\% centralities from IQMD. For the odd-order collective flow, the protons and deuterons are selected within rapidity of -0.5 - 0, while for the even-order collective flow, the rapidity is selected within -0.1 - 0. The transverse momentum is selected within 1-1.5 GeV/c.
      }
      \label{fig6}
      \end{figure}

Fig.~\ref{fig6} shows the dependence of the $\varepsilon_{n}$ and $\left \langle v_{n} \right \rangle$ as well as the ratio $\left \langle v_{n} \right \rangle/\varepsilon_{n}$ on the centrality. It is obvious that the collision  eccentricity increases as centrality (here larger centrality corresponds to more peripheral collision), indicating a more elliptical structure under more off-central collisions. The $\left \langle v_{2} \right \rangle$ and $\left \langle v_{2} \right \rangle/\varepsilon_{2}$ is negative, and decrease with the increase of centrality. The large positive elliptic flow in RHIC energy region (above 3 GeV) in semi-central collisions are always understood by the hydrodynamic picture. With the dynamic evolution of the fireball~\cite{lacey2011initial,alver2010collision,de2012effects}, the geometric anisotropy of the initial state will be transformed into the anisotropy of the momentum space at final state which is characterized by the collective flow, $\left \langle v_{n} \right \rangle$. However at lower energy the large negative elliptic flow in non-central collisions was explained by the squeeze-out mechanism~\cite{squeeze,Wang_PRC,WangTT_EPJA}. From Fig.~\ref{fig6}, it is seen that the negative elliptic flow presents larger absolute value in non-central collisions where the $\varepsilon_2$ also takes  larger value than that in central collisions. This indicates that more spectators in non-central collisions enhance the particle emission out of plane by the squeeze-out mechanism.

The initial fluctuation results in the triangular and higher-order asymmetry of the geometry shape in the reaction zone. The third ($\varepsilon_{3}$) and fourth ($\varepsilon_{4}$) order eccentricity coefficients are calculated for the reaction system by using the participants, as shown in  Fig.~\ref{fig6} (b) and (c).  As $\varepsilon_{2}$, both $\varepsilon_{3}$ and $\varepsilon_{4}$ increase with centrality, which implies that the initial fluctuation is more obvious in peripheral collisions than in central collisions. The initial geometry fluctuation determines the high-order harmonic flows in momentum space and the centrality dependent trend, which is shown in Fig.~\ref{fig6}(e) and (f). $\left \langle v_{3} \right \rangle$ present the similar centrality dependence as $\varepsilon_{3}$ which is consistent with the phenomena in ultra-relativistic heavy-ion collisions~\cite{alver2010collision}. $v_{4}$ weakly depends on centrality and the nonlinear-mode is not separated~\cite{ZHANG2020135366-PLB2020}, which is beyond the scope for this work. The ratios of $\left \langle v_{n} \right \rangle/\varepsilon_{n}$ ($n=3, 4$) as shown in Fig.~\ref{fig6}(h) and (i) also indicates that there is more significant initial fluctuation effects  in peripheral collisions than in central collisions.
      
      \par

      \subsection{Linear-correlation between collective flows}

  \begin{figure}[htb]
      \includegraphics[angle=0,scale=0.23]{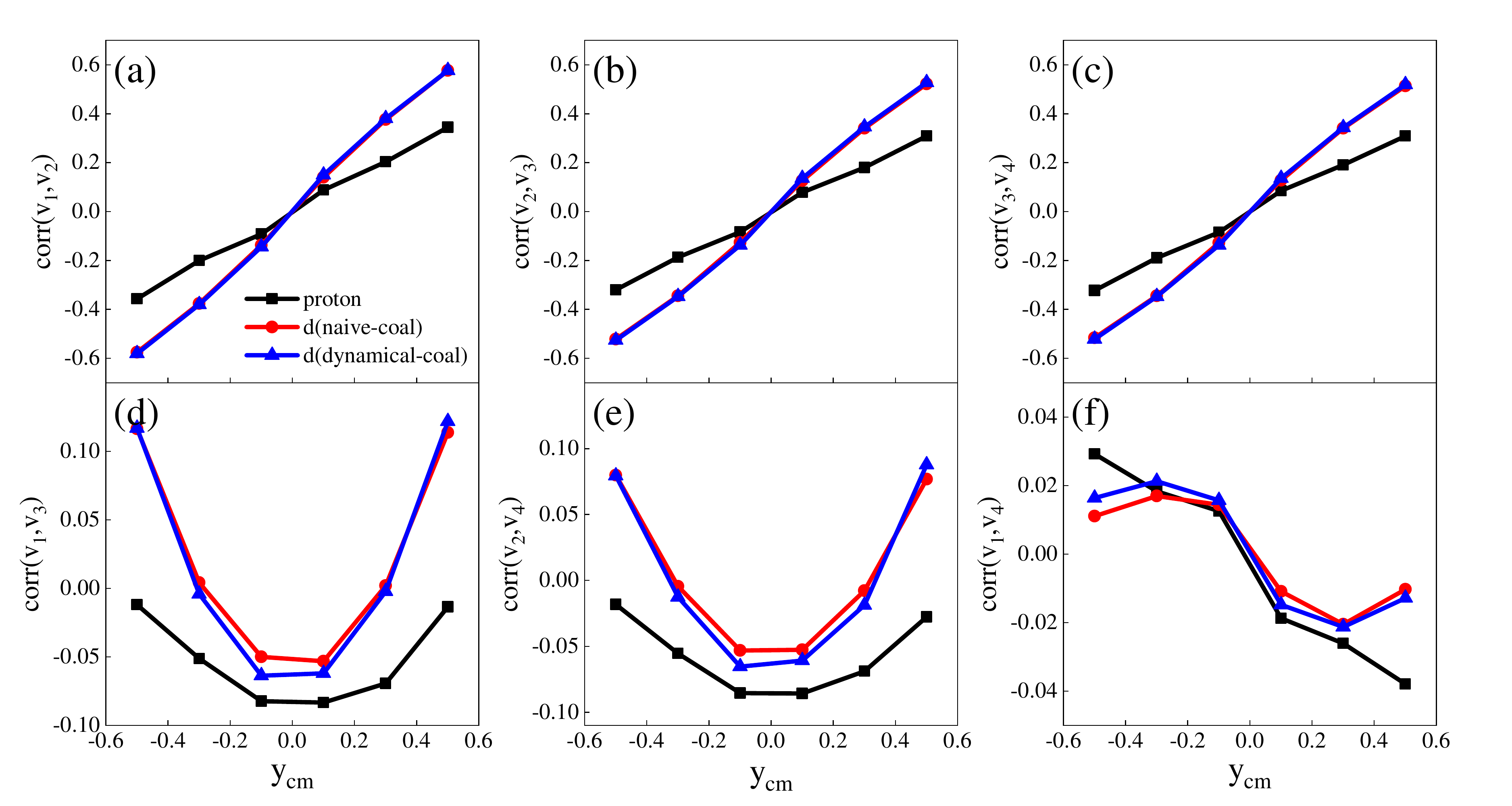}
      \caption{The Pearson correlation function $corr(v_{n},v_{m})$ of protons and deuterons as a function of rapidity in Au + Au collisions at 1.23A GeV from IQMD. The transverse momentum of protons and deuterons are selected as 1-1.5 GeV/c.
      }
      \label{fig7}
      \end{figure}  
      
      \begin{figure}[htb]
      \includegraphics[angle=0,scale=0.23]{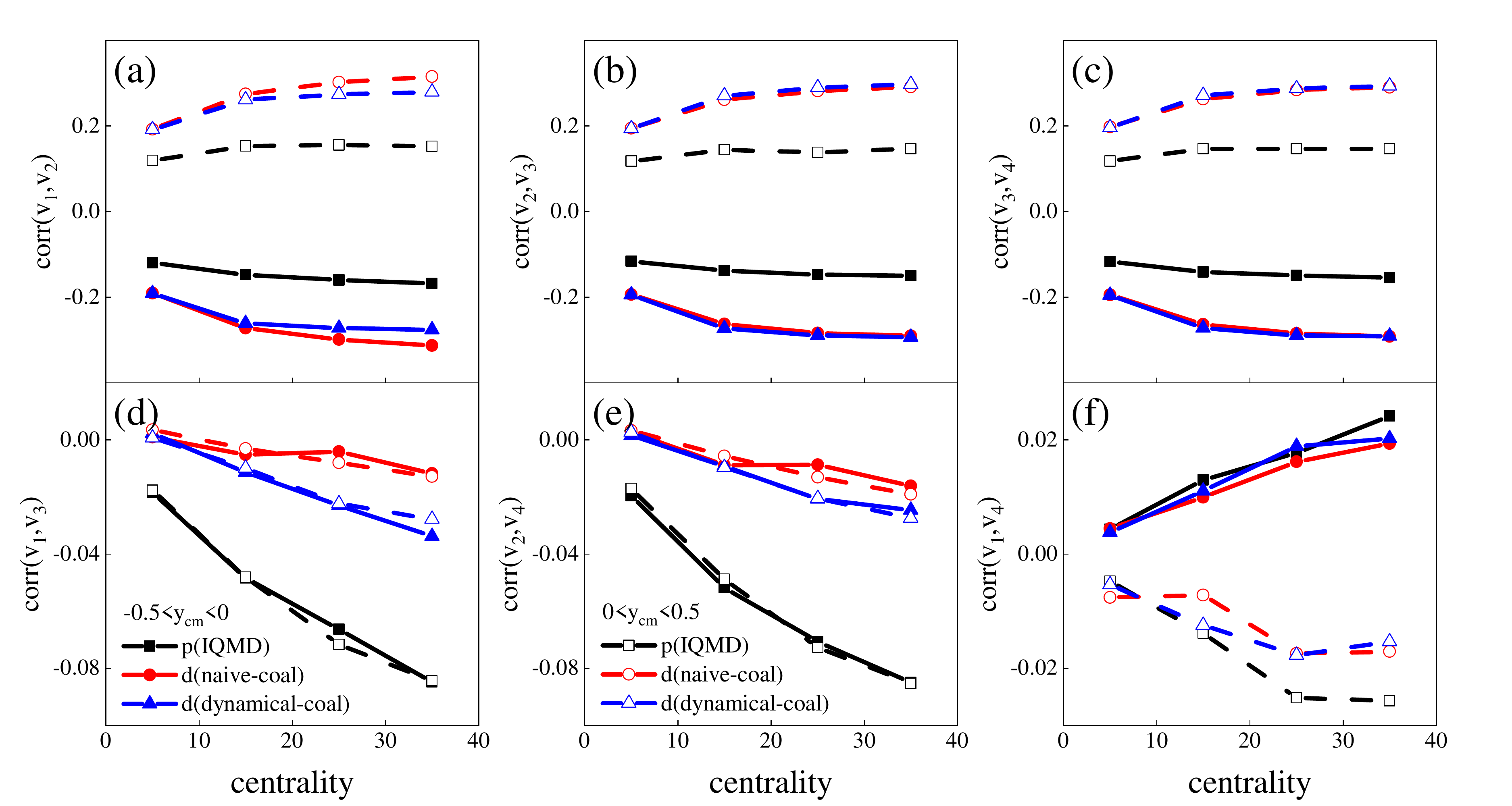}
      \caption{The Pearson correlation function $corr(v_{n},v_{m})$ of protons and deuterons as  a function of centrality in Au + Au collisions at 1.23A GeV from IQMD. The transverse momentum of protons and deuterons are selected as 1-1.5 GeV/c. The open symbols represent for $ 0 < y_{cm} < 0.5$ and the solid symbols represent for  $ -0.5 < y_{cm} < 0$.
      }
      \label{fig8}
      \end{figure}
      
      To further understand the coalescence mechanism of deuteron in the collisions, the linear correlation functions $corr({v_{n},v_{m}})$ known as the Pearson coefficient for protons and deuterons are calculated as~\cite{reichert2022harmonic}:
      
      \begin{eqnarray}
	 \begin{aligned}
      corr\left ( v_{n},v_{m} \right )=\frac{\left \langle v_{n}v_{m} \right \rangle-\left \langle v_{n} \right \rangle\left \langle v_{m} \right \rangle}{\sigma _{v_{n}}\sigma _{v_{m}}}.
      \label{eq:corr_vn}
       \end{aligned}
      \end{eqnarray}
Here, the standard deviation $\sigma _{v_{i}} = \sqrt{\left \langle v_{i}^{2} \right \rangle-\left \langle v_{i} \right \rangle^{2}}$ is used to normalize the covariance. We know that the Pearson coefficient provides a measure for linear dependence of two random variables, which equals to 1 implies a perfect linear dependence, but a vanishing Pearson coefficient does not rule out any nonlinear correlation.

We show the Pearson correlation function $corr(v_{n},v_{m})$ between the first four flow coefficients of protons and deuterons in Au+Au collisions at 1.23 AGeV from IQMD modle as a function of rapidity in Fig.~\ref{fig7}, and as a function of centrality in Fig.~\ref{fig8}.

In Fig.~\ref{fig7}, we can see that  the correlation between the even and odd flow harmonic, for example, $corr(v_{1},v_{2})$ (Fig.~\ref{fig7}(a)),  $corr(v_{2},v_{3})$ (Fig.~\ref{fig7}(b)) and $corr(v_{3},v_{4})$ (Fig.~\ref{fig7}(c)) as well as $corr(v_{1},v_{4})$ (Fig.~\ref{fig7}(f)), are antisymmetric, and while the correlation between the even/odd flow harmonic, for example, $corr(v_{1},v_{3})$ (Fig.~\ref{fig7}(d)) and $corr(v_{2},v_{4})$ (Fig.~\ref{fig7}(e)), are symmetric around $y_{cm}=0$. This phenomenon is consistent with the conclusion given in Ref.~\cite{reichert2022harmonic}. Moreover, we can see that the correlation between adjacent-order $v_{n}$, for example, $corr(v_{1},v_{2})$ (Fig.~\ref{fig7}(a)), $corr(v_{2},v_{3})$ (Fig.~\ref{fig7}(b)) and $corr(v_{3},v_{4})$ (Fig.~\ref{fig7}(c)), is stronger, and the results are basically the same. Furthermore, we can find an interesting phenomenon that the correlations between different order $v_{n}$ is closely related to the differentials between the order numbers. As differential between orders equals to 1, such as $corr(v_{1},v_{2})$ (Fig.~\ref{fig7}(a)), $corr(v_{2},v_{3})$ (Fig.~\ref{fig7}(b)) and $corr(v_{3},v_{4})$ (Fig.~\ref{fig7}(c)), has the same result. As differential between orders equals to 2, such as $corr(v_{1},v_{3})$ (Fig.~\ref{fig7}(d) and $corr(v_{2},v_{4})$ (Fig.~\ref{fig7}(e)), has the similar result. Moreover, the smaller the differential between the order numbers, the larger the correlation between $v_{n}$, which indicates that the correlation between $v_{n}$ of the more adjacent order is stronger. 

 Considering the (anti-)symmetry behavior of flow correlation coefficients as a function of rapidity, we extract the average $corr(v_{n},v_{m})$ in positive and negative rapidity intervals, which is shown in Fig.~\ref{fig8}. We observe that as the centrality increasing, the Pearson coefficient between different order $v_{n}$ increases gradually, but the increasing trend is somewhat different. Compared with the obvious increase of the correlation between the first and second (Fig.~\ref{fig8}(a)), the second and third (Fig.~\ref{fig8}(b)), and the third and fourth (Fig.~\ref{fig8}(c)) flow harmonic, we notice the correlation between the first and third (Fig.~\ref{fig8}(d)), the second and fourth flow (Fig.~\ref{fig8}(e)), and the first and fourth (Fig.~\ref{fig8}(f)) harmonic increases slightly which can be ignored overall. Overall, the above  correlation phenomenon is interesting, but the deeper understanding needs to be further studied in the future.

      \section{Summary}
      \label{sec:summary}
      \par
In summary, the yields of protons and deuterons were calculated by a simulation of Au + Au collision at 1.23A GeV with the IQMD model and coalescence models. Then by an event-plane method, we calculate the first four order collective flows of protons and deuterons. The results show that a good nucleon-number scaling of elliptic flow among proton and deuteron holds. The ratio $\left \langle v_{4} \right \rangle/\left \langle v_{2} \right \rangle^{2}$ approaches to the experimental value of 1/2 with a large error between $\pm$0.3 rapidity but decreases beyond mid-rapidity interval, and $\left \langle v_{3} \right \rangle/(\left \langle v_{1} \right \rangle \left \langle v_{2} \right \rangle)$ is higher than those from the HADES experiment. In addition, we give the dependence of $\varepsilon_{n}$, $\left \langle v_{n} \right \rangle$ as well as $\left \langle v_{n} \right \rangle/\varepsilon_{n}$ ratio on the centrality, indicating a more elliptical structure under more off-central collisions, and more spectators in non-central collisions enhance particle emission out of plane by the squeeze-out mechanism. At last, we show the rapidity and centrality dependence of the linear correlation coefficients $corr(v_{n},v_{m})$ between the first four flow coefficients, 
and  notice that the correlation between the even/odd flow harmonic are symmetric around $y_{cm}=0$, while the correlation between the even and odd flow harmonic are antisymmetric. The correlations between different order $v_{n}$ is closely related to the differentials between the order numbers. For the Pearson correlation functions $corr(v_{n},v_{m})$ with the same differentials have the same result. And the smaller the differential between the order numbers, the larger the correlation between $v_{n}$, which indicates that the correlation between $v_{n}$ of the more adjacent order is stronger.  From  the centrality dependence,  the Pearson coefficient between different order $v_{n}$ increases gradually. Further understanding of physics insight to different harmonic flow correlation is expected.

      \begin{acknowledgements}
      This work was supported in part by the National Natural Science Foundation of China under contract Nos. 11890714, 11875066, 11421505, 11775288, and 12147101, the National Key R\&D Program of China under Grant Nos. 2016YFE0100900 and 2018YFE0104600, and by Guangdong Major Project of Basic and Applied Basic Research No. 2020B0301030008.
      \end{acknowledgements}

      \end{CJK*}

      \bibliography{myref}

      \end{document}